\documentclass[apjl]{emulateapj}
\usepackage{natbib}
\usepackage{subfigure}

\shorttitle{The Trailing Edge of the KH~15D Circumbinary Ring}
\shortauthors{Capelo et~al.}
\begin{document}

\title{Locating the Trailing Edge of the Circumbinary Ring in the KH~15D System}

\author{Holly L. Capelo\footnote{Current address: Max Planck Institute for Dynamics and Self-Organization, Am Fa\ss burg 17, 37077 G\"{o}ttingen, Germany} and William Herbst}
\affil{Astronomy Department, Wesleyan University, Middletown, CT 06459}

\author{S. K. Leggett}
\affil{Gemini Observatory, Northern Operations Center, 670 N. A'ohoku Place, Hilo, HI 96720}

\author{Catrina M. Hamilton}
\affil{Department of Physics and Astronomy, Dickinson College, Carlisle, PA 17013}

\author{John A. Johnson}
\affil{Department of Astrophysics, California Institute of Technology, 1200 E. California Blvd., Pasadena, CA 91125}

\begin{abstract}
Following two years of complete occultation of both stars by its opaque circumbinary ring, the binary T Tauri star KH~15D has abruptly brightened again during apastron phases, reaching $I = 15$ mag. Here, we show that the brightening is accompanied by a change in spectral class from K6/K7 (the spectral class of star~A) to $\sim$K1, and a bluing of the system in $V-I$ by about 0.3 mag. A radial velocity measurement confirms that, at apastron, we are now seeing direct light from star~B, which is more luminous and of earlier spectral class than star~A. Evidently, the trailing edge of the occulting screen has just become tangent to one anse of star~B's projected orbit. This confirms a prediction of the precession models, supports the view that the tilted ring is self-gravitating, and ushers in a new era of the system's evolution that should be accompanied by the same kind of dramatic phenomena observed from 1995-2009. It also promotes KH~15D from a single-lined to a double-lined eclipsing binary, greatly enhancing its value for testing pre-main sequence models. The results of our study strengthen the case for truncation of the outer ring at around 4~AU by a sub-stellar object such as an extremely young giant planet. The system is currently at an optimal configuration for detecting the putative planet and we urge expedient follow-up observations.
 \end{abstract}

\section{Introduction}

\objectname{KH~15D}  is a young binary system composed of similar, but not identical, low-mass pre-main sequence stars in an orbit of eccentricity $\sim$0.6 with a period of 48.37 days \citep{johnson04,winn04,hamilton05}. The binary orbit is viewed nearly edge-on and is embedded in an accretion disk from which a well-collimated outflow emerges \citep{hamilton03,deming04,tokunaga04,mundt10}. The estimated age of the stars is $3\times 10^{6}$ yr, based on their location in the color-magnitude diagram of NGC 2264, and the total system mass is $\sim$1.3 M$_\odot$ \citep{hamilton01,johnson04}. A thin circumbinary ring of evolved solids has precipitated from the gas disk \citep{lawler10} within the terrestrial zone (1-4~AU) of this system \citep{herbst08}. The ring reveals itself as an opaque screen with a razor-sharp edge ($\lesssim$ 0.1 stellar radii; \citeauthor{herbst10} 2010) at optical and near-infrared wavelengths. For decades, the leading edge of this structure has been slowly moving across the binary orbit, apparently resulting from precession of the slightly inclined ring \citep{chiang04,winn04,winn06}. One star's orbit (designated star~B) was completely occulted in 1995 and the other (star~A) in 2009 \citep{herbst10}. 

The tidal influence of a binary on a misaligned disk is known to cause rigid-body precession given that the disk can communicate at length-scales comparable to the inner disk radius and time-scales close to the local precession period \citep{arty94,pap95,larwood97}. The variable torque applied by a non-coplanar binary will truncate the inner disk radius and induce a gradient of inclinations across concentric annuli in the disk, producing a warp. The specific geometry of the warp, such as whether the gradient increases or decreases with radius, depends upon the mechanism responsible for enforcing an alignment of the nodes of the annuli. \cite{chiang04} propose two means by which the KH~15D circumbinary ring maintains such an alignment - either by self-gravity of the ring particles or by thermal pressure normal to the ring plane - with a unique warp geometry resulting in each case.  In such warped, precessing disks, if the outer radius is allowed to become arbitrarily large, then the disk precession period will approach infinity. Thus, for nodal precession to occur, the outer disk must also be confined in radial extent (i.e. it forms a small, separate ring), such as by a shepherding body \citep{ goldreich82, chiang04} or a sufficiently rapid decrease in surface density \citep{larwood97}. The misalignment of the binary orbital plane with the KH~15D ring is believed to induce precession in its ring and, further, the system exhibits distinctive eclipses due to this non-coplanar arrangement.  

Although identified as a variable star in the late 1960s \citep{badalian70}, interest in KH~15D began with the discovery in 1995 \citep{kearns98} that the star essentially winked on and off by several magnitudes on a period of 48.37 days \citep{hamilton05}, now known to be the orbital period of the binary \citep{johnson04}. These extreme brightness excursions started then because precession of the ring had moved one of its razor-sharp edges fully across the orbit of one member of the binary (star~B), leaving the other (star~A) to periodically rise and set with respect to that horizon as seen from Earth. From 1995 to 2009, the winking continued, with the system being increasingly more ``off'' as star~A spent more and more of each orbital cycle below the horizon \citep{hamilton05,herbst10}. In 2010 and 2011, the system was fainter at all times because neither star appeared above either edge. 

\begin{figure}[Ht]
\centering
\includegraphics[scale=.55,trim=1.4cm 0.5cm 0.9cm 0.6cm, clip=true]{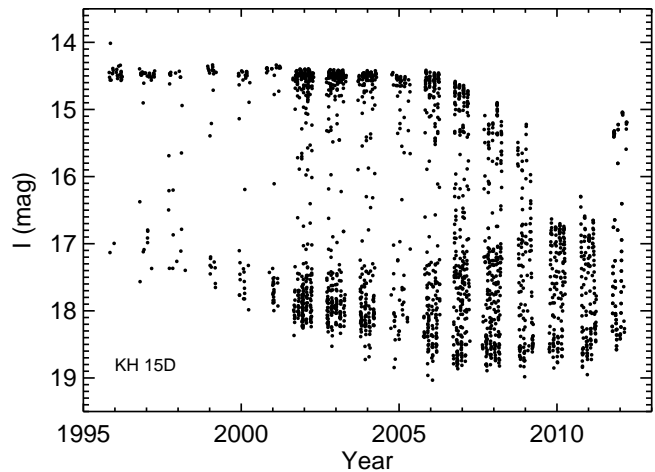}
\caption{\label{longterm}The long-term light curve of KH~15D showing its dramatic re-brightening in 2012. The one measurement in 1995 showing the system at $I=14$ mag was obtained near periastron, when star~B made its last appearance above the leading edge of the screen.}
\end{figure}

Both stars are believed to be less massive than the Sun, with star~A having a spectral type of K6/K7 \citep{hamilton01,agol04}. The prediction that star~B is slightly more massive and of earlier type has heretofore only been inferred from the fact that it is a brighter star, as evidenced by archival photometry \citep{winn03,johnsonwinn04,johnson05} and by its one appearance in the modern photometric record (see Fig.~\ref{longterm}) prior to 2012. 

\section{Observations and Data Analysis}

We obtained images of KH~15D on over three hundred nights between October 2010 and March 2012, using the ANDICAM instrument on the 1.3m telescope operated by the SMARTS consortium at Cerro Tololo Inter-American Observatory (CTIO). The $I-$band photometry is in addition to an observing record from multiple observatories dating back to 1995 and the $V-I$ color information has been on record for nearly a decade. The optical images consisted of four exposures, each lasting 150 seconds. We used IRAF scripts to flat field, sky subtract and combine dithered images, and the package \textit{phot} to complete the aperture photometry. The median error is 0.01 mag for $I  < 17.5$ mag and increases to 0.05 mag at $I=18.5$ mag.

We obtained cross-dispersed (XD) spectroscopy with the Gemini Near Infrared Spectrograph (GNIRS) on UT January 5, 2011, when the stars were both near apastron but fully obscured, with $I=16.95$ mag. The XD data were taken in the GNIRS short focal length camera mode with wavelength coverage from  0.9 to 2.5 $\mu$m, a slit width of 0.3$''$, and a 31.7 line mm$^{-1}$ grating (1600 $\lesssim$ R $\lesssim$ 1700). The processing used the  \textit{gemini.gnirs} IRAF package Beta Version 1.11. The data products consist of a set of calibration images, a spectrum of KH~15D and a spectrum of HD 59374, an F8 V star observed for the purposes of characterizing the sensitivity function. The integration times were less than four seconds for all of the calibration and standard exposures. The observations of the target itself include 20 frames, each of 300-second duration. The spectra were flat-fielded, assigned a wavelength solution by comparison with an argon arc-lamp spectrum and corrected for spatial distortions and detector artifacts. The details of the standard GNIRS reduction procedures are given in \cite{winge09}. 

\begin{figure}[Ht]
\includegraphics[scale=.54,trim=1.4cm 0.5cm 0.55cm 0.6cm, clip=true]{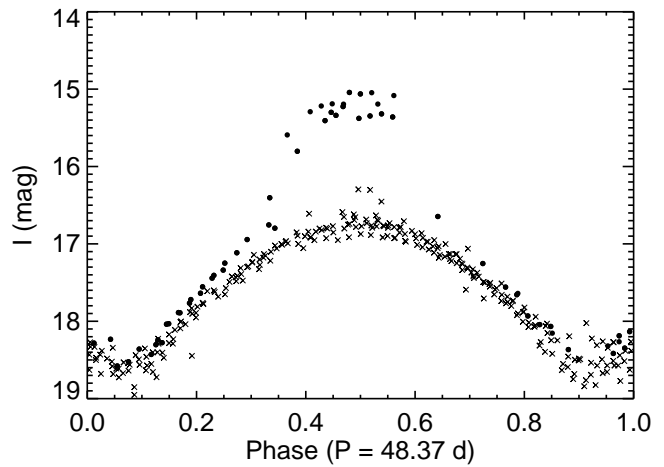}
\caption{\label{last3lightcurves} Light curves from 2010 and 2011 (x's) and 2012 (solid circles) phased with the orbital period of 48.37 days. The dramatic brightening that occurred around apastron (phase $=0.5$) in 2012 is apparent and represents the appearance of star~B partially above the ring edge. }
\end{figure}

\section{Results} \label{results}

\begin{figure*}[Ht]
\centering
\includegraphics[scale=1.0,trim=0.0cm 0.0cm 0.0cm 0.0cm, clip=true]{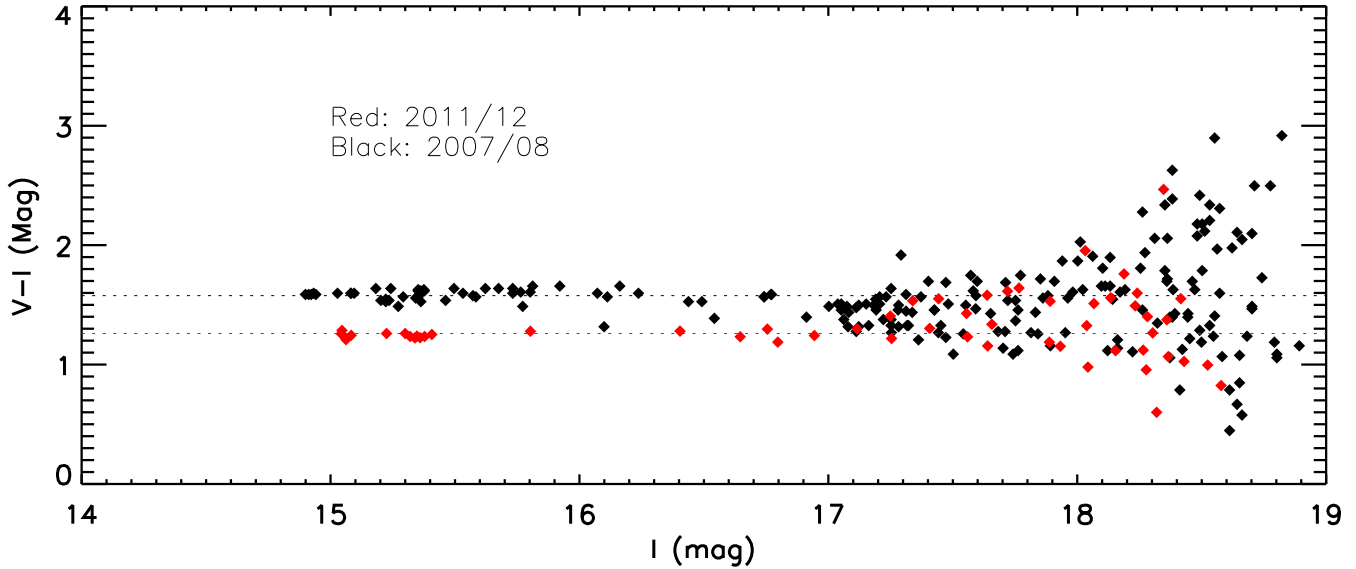}
\caption{\label{colorevol} A comparison of the colors in 2012 to four years prior. The set of points which collect around a relatively blue median color of $V-I = 1.25$ mag represent star~B and those that collect around a redder median color of $V-I = 1.56$ mag represent star~A. }
\end{figure*}

\subsection{Lightcurve}\label{lightcurves}
  The $I-$band light curve of KH~15D from 1995 to present is shown in Fig.~\ref{longterm}.
Each vertical strip represents an annual observing season, culminating in October-March 2012. The full on-off eclipse behavior, caused by star~A disappearing behind the leading edge of the occulting screen during each orbit, was present up until about 2006. The maximum system brightness then gradually diminished, because star~A never fully rose above the screen edge. The system spent increasingly more time in complete obscuration until 2010, when both stars were fully occulted at all phases and the system never appeared brighter than $I  = 16.3$  mag. Interestingly, even when fully occulted, the system varied by more than two magnitudes on the orbital period, being brighter near apastron and fainter near periastron, due to scattered light (see Fig.~\ref{last3lightcurves}).

The re-brightening of the system that occurred in 2012, reaching $I = 15$ mag near apastron (phase $= 0.5$), is apparent on Fig.~\ref{longterm} and highlighted on Fig.~\ref{last3lightcurves}, where we show light curves for each of the past three seasons, phased to the orbital period.  A ``star rise" is evident during one cycle in 2012 at a phase of 0.35 and a ``star set" during a different cycle is seen at phase 0.65. The maximum brightness near $I = 15$ mag is less than the known brightness of star~A ($I = 14.5$ mag) or star~B ($I = 14.0$ mag) and, therefore, most likely represents a star still partially occulted by the ring. This is analogous to the lightcurve in the 2008 season, when star~A was partially occulted. We now show that this is not star~A reappearing, but star~B appearing from behind the trailing edge of the occulting screen.

Fig.~\ref{colorevol} shows the $V-I$ color as a function of brightness, comparing the photometry from 2008 and 2012. The horizontal lines in the figure show the median color for each season when the system is bright, i.e., after star rise. In 2008, we found $V-I=1.56$ mag, characteristic of the K6/K7 star (A). In 2012, we measure $V-I=1.25$ mag, significantly different and characteristic of a hotter star. Correcting for foreground reddening [$E$($B-V$) $= 0.15$ \citep{hamilton01}, corresponding to $E$($V-I$)$ = 0.21$], we find the intrinsic color of the (partly) visible star in 2012 to be ($V-I$)$_{0} = 1.04$ mag, characteristic of a K1 giant \citep{Bessell79}. Note that there is no evidence of any reddening accompanying the large extinction suffered by the stars, either in 2012, or in any year between 1995 and 2009, when star~A was undergoing regular rise and set events.

\subsection{Spectral Classification}\label{classification}

\begin{figure*}[Ht]
 \begin{centering}
\includegraphics[angle=90,scale=0.7]{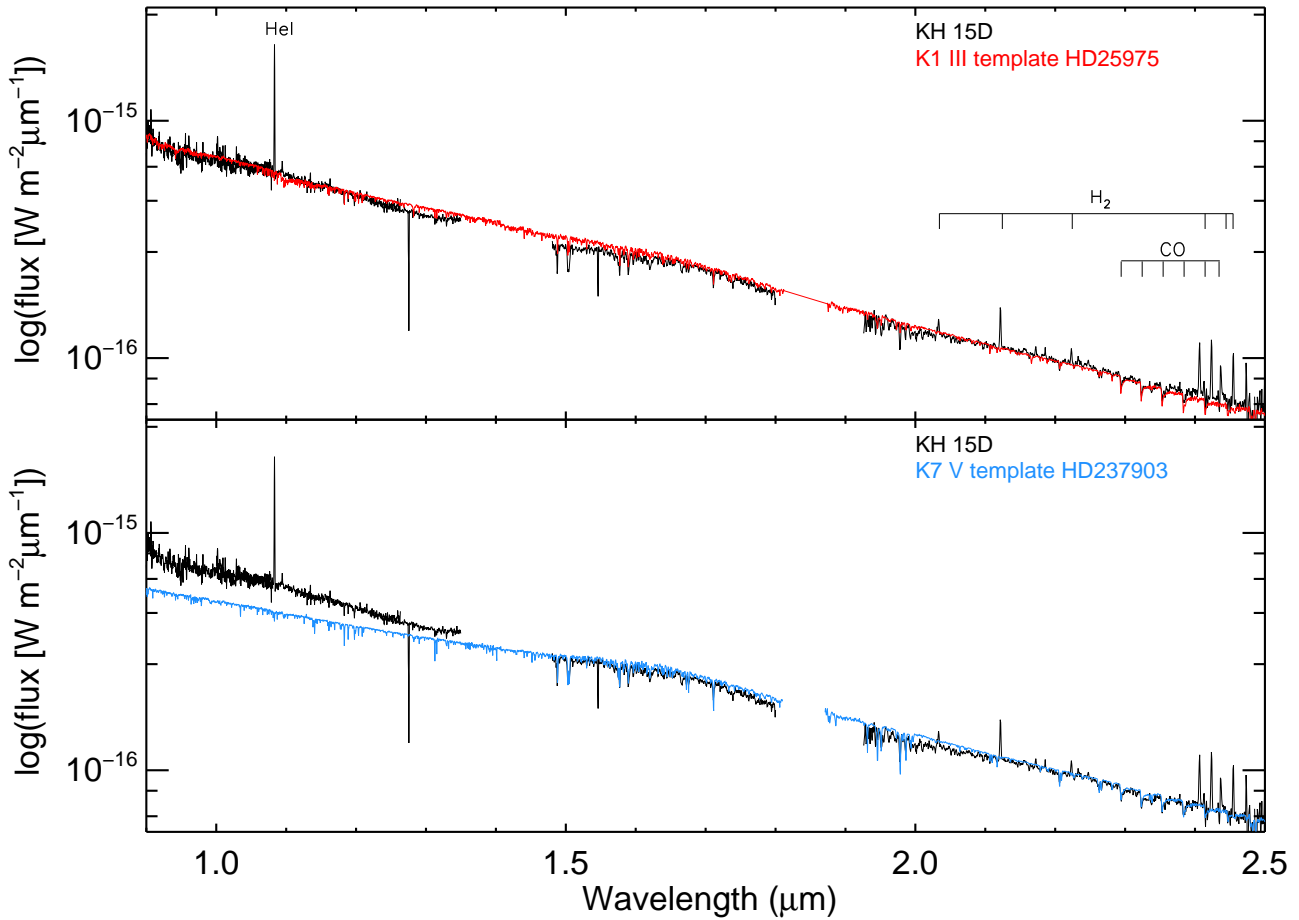}
\caption{\label{spectra}The GNIRS spectrum of KH~15D (in black) compared to template spectra of type K1~III (top, red) and K7~V (bottom, blue). Some of the absorption features unique to the KH 15D spectrum, including the prominent ones at 1.275 and 1.546 $\mu$m, are as yet unidentified.}
\end{centering}
\end{figure*}

The Gemini/GNIRS spectrum obtained in January 2011, when the system had $I=16.95$ mag, supports our photometric result that a bluer star, presumably star~B, recently became a significant contributor to the light of KH~15D. Fig.~\ref{spectra} shows the spectrum spanning five of the XD orders (0.9 to 2.5  $\mu$m).  Obvious features are labeled; the molecular hydrogen emission features most likely arise from the shocked ambient medium \citep{deming04,kusakabe05}, the HeI line from the funnel accretion flow \citep{Fischer08}, and the CO band heads from the stellar photosphere itself.   The spectrum was flux calibrated by integrating over the 2MASS filter profiles and scaling the observations to fit.  The blue end was calibrated using the overlap region with the $J$ band. The source variability introduces error into the flux normalization, but does not affect the spectrum's shape nor relative line strengths, which we now use to classify the source by visually comparing it to template spectra \citep{pickles, spex1, spex2}. In Fig.~\ref{spectra}, KH~15D  is in black and the red (top) and blue (bottom) spectral templates correspond to HD 25975, a K1~III star, and to HD 237903, a K7~V star, respectively. The continuum slope is inconsistent with a spectral class as late as K6/7, as expected of star~A: the KH~15D and K7~V star continua cease to overlap blueward of 1.3 $\mu$m. Matching the continuum slope of the full wavelength region in question requires a contribution from a star of type K1$\pm$0.5. The set of lines between 1.1 and 1.2 $\mu$m and also those just before 2 $\mu$m would be deeper, were the K7 star the only source.  This spectrum was obtained near apastron but at a time when both stars were fully obscured so it represents the combined scattered light. We note that since the stars interact with one another and their disk (see e.g. \citeauthor{Hamilton12} 2012), they might defy a clear luminosity classification, but as they are still contracting to the main sequence, one expects them to effectively be sub-giants or giants, in agreement with our findings. 

\subsection{Radial Velocity}
To confirm our interpretation that it is star~B that has caused the system to brighten, we obtained a 10-minute exposure of KH~15D with the Keck High Resolution Echelle Spectrometer (HIRES) on UT March 13, 2012, when the system was near its maximum brightness.  The radial velocity (RV) was measured by cross-correlation with a K0 main sequence star (HD 185144) yielding a heliocentric value of +39.9 km s$^{-1}$.  At this orbital phase (0.41), the heliocentric velocity of star~A is predicted to be -0.1 km s$^{-1}$ (Model 3, \citeauthor{winn06} 2006) or +2 km s$^{-1}$ \citep{johnson04}.  We caution the reader that the systemic velocity of the KH~15D system is not yet well determined due to the difficulty in obtaining spectra suitable for cross-correlation analysis while deep in eclipse.  However, two values exist in the literature (+14.7 km s$^{-1}$, \citeauthor{johnson04} 2004 and +18.6 km s$^{-1}$, \citeauthor{winn06} 2006), which are consistent with the radial velocities of the stars in the NGC 2264 cluster \citep{Soderblom99}. We adopt the value of \cite{winn06}, finding that relative to the center of mass, star~B is moving with a velocity of $+21.3\pm1$ km s$^{-1}$, while the predicted velocity for star~A is $-18.7\pm1$ km s$^{-1}$. The uncertainty represents the standard deviation of the mean RV measured from all echelle orders. These velocities suggest that the two stars are of similar mass, however, that star~A, not star~B, is the more massive of the pair.  While this seems contradictory (as the brighter, hotter star~B should be the more massive one), we urge discretion regarding this point since the value used for star~A is model-dependent. Clearly, once more RV measurements are obtained as the orbit of star~B is revealed, the models can be refined.  More importantly, the RV measurement confirms that we are now observing star~B at apastron, not star~A.

\section{Discussion}
\subsection{Confirmation of the Precession Model}

The precessing circumbinary-ring model has provided a successful framework for interpreting observations of KH~15D \citep{winn04,winn06,chiang04,silvia08}. A prediction is that the eclipse behavior exhibited by star~A from the early 1990s through 2009 should repeat itself with star~B as the visible star. The developments reported here confirm that prediction and provide strong support for this scenario. Fig.~\ref{screenmotion} is a schematic drawing of the system, as we believe it appears in 2012. For the first time, we are able to locate the trailing edge of the ring which provides important new information to improve the next generation of models.

\cite{winn04,winn06} employed archival records to address the secular evolution of KH~15D by projecting the binary orbit onto the plane of the sky and treating the precession of the ring as a sharp, semi-infinite edge (that is, the location of leading edge is finite and the trailing edge extends to infinity) that marches across the projected binary orbit over time. The authors predicted the eventual re-emergence of star~B as a natural result of the advancing edge. As the model had no way to differentiate between viewing angles with respect to the occulting feature nor its thickness, it provided no definite prediction regarding whether both binary orbits would ever be occulted at the same time or, if so, when star~B  might reappear.  Having the location of only one edge of the screen specified also complicated the modeling of the scattered light. The discovery of the location of the trailing edge should lead to a much improved version of this model which will ultimately quantify most of the important physical characteristics of the system.

\cite{chiang04} presented a first-order dynamical theory of the KH~15D ring noting that its rigid precession can be enforced either by thermal pressure gradients or by ring self-gravity. They favored self-gravity and their case is strengthened by the observations contained herein. The future light curve of KH 15D predicted by \cite{chiang04} for the thermal pressure case (Model~1) differed substantially from that for self-gravity (Model~2) because the direction of warping is different. In Model~1, the disk is closer to the plane at large distances and more inclined closer in and the predicted duration of full occultation is quite long (many decades). In Model~2, the disk is more inclined at larger distances and the duration of full occultation of both orbits is short or non-existent. We now know that the observed light curve had a short duration ($\sim$2 years) of full occultation, qualitatively in much better agreement with the self-gravity model, although quantitatively in agreement with neither.

While recent data will enable properties such as the ring geometry, orientation and scattering properties to be refined, the mounting support for rigid nodal precession underscores that the occulting ring of solids must be confined at its outer borders. Therefore, the inferred radial dimensions of the ring should not change drastically from past estimates. 

\subsection{Remaining Questions and Future Observations}

The question remains as to what has truncated the ring at around 4~AU. \citet{chiang04} suggest a planet as the most likely cause. Being at or near the ice line in the system it is possibly a giant planet in formation. Color excursions to values redder than star~A have long been known in KH~15D during minimum light and may find their interpretation in the light of this putative third body.  Such an extremely young giant planet, if sufficiently massive, might be self-luminous enough to be detectable in the near or mid-infrared during phases when both stars are occulted. It should be sought immediately, as  the minimum brightness of the system is already beginning to gradually increase due to the advancement of the trailing edge (see Fig.~\ref{longterm}).  

Fig.~\ref{last3lightcurves} shows that the abrupt change in slope that occurs during star rise and star set continues, suggesting that the trailing edge of the ring will be as sharp as the leading edge: namely, much smaller than a stellar radius \citep{herbst10}. Resultantly, the eclipses that will ensue will allow us to determine the radius of star~B with the same precision that can be achieved in a typical eclipsing binary. We can now obtain the RV curve of star~B, and so will constrain the masses, radii, luminosities and effective temperatures of the stars. KH~15D will soon be the first known double-lined eclipsing binary young enough to still be embedded in an accretion disk, making it of paramount importance for testing models of pre-main sequence evolution. 

\begin{figure}[Ht]
\begin{centering}
\includegraphics[scale=.43,angle=90,trim=5.0cm 0.3cm 3.0cm 2.5cm, clip=true]{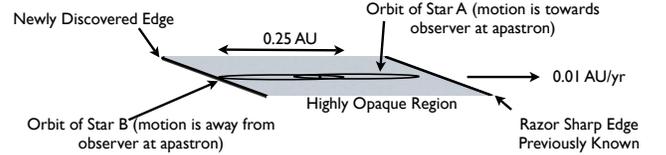}
\caption{\label{screenmotion} A schematic representation of the KH~15D system as projected on the sky.}
\end{centering}
\end{figure}

As Fig.~\ref{screenmotion} shows, the physical extent of the obscuring portion of the ring is slightly wider than the projected size of the binary orbits. What sets this width remains unclear to us. Multiple possibilities exist but it is well beyond the purview of this observational paper to assess them here. The KH~15D spectral features of non-stellar origin will be addressed in subsequent work, and additional spectra should be obtained to verify these features and probe more sight lines in the ring.  Over the coming decade we will have the opportunity to use the trailing edge for natural coronography to study the magnetosphere and jet-launching zone of star~B in the same detail as was done for star~A.

\section{Summary}

The reappearance of star~B provides a dramatic confirmation of the precessing-ring model, demonstrating that such structures must be a common feature of circumbinary and, presumably, circumstellar transition disks. It opens the door for a more complete study of the ring and what its role in planet formation may be. This is now more pertinent than ever, as the Kepler mission continues to discover circumbinary planetary systems \citep{Doyle11,2012Natur.481..475W}. Follow-up observations of KH~15D should be conducted forthwith to uncover the putative planet confining the circumbinary ring. The emergence of star~B has transformed KH~15D from a single-lined spectroscopic binary to a double-lined eclipsing system, with all that that entails for precise measurement of fundamental astrophysical properties. 

\acknowledgments{Acknowledgments}

We thank Geoff Marcy for helping to arrange the Keck/HIRES observations. We are grateful to the observers and personnel at the SMARTS and GNIRS facilities. This work was partially supported by a NASA grant to W.H. through the Origins of Solar Systems program. Based on observations obtained at the Gemini Observatory via queue program GN-2010B-Q-52. Gemini is operated by the Association of Universities for Research in Astronomy, Inc., under an agreement with the NSF on behalf of the Gemini partnership: the National Science Foundation (U.S.), the Science and Technology Facilities Council (U.K.), the National Research Council (Canada), CONICYT (Chile), the Australian Research Council (Australia), Minist\'{e}rio da Ci\^{e}ncia, Tecnologia e Inova\c{c}\~{a}o (Brazil) and Ministerio de Ciencia, and Tecnolog\'{i}a e Innovaci\'{o}n Productiva (Argentina).

\bibliographystyle{apj}

\clearpage
\end{document}